\documentclass[english,letterpaper,twocolumn,showpacs,pra,aps]{revtex4}
\usepackage{graphicx}
\usepackage{amssymb}

\makeatletter

\large  

\input epsf

\makeatother

\usepackage{babel}
\makeatother
\begin{document}

\title{Weyl problem and Casimir effects in spherical shell geometry}

\author{Eugene B. Kolomeisky$^{1}$, Hussain Zaidi$^{1}$, Luke Langsjoen$^{1}$, and Joseph P. Straley$^{2}$}

\affiliation{$^{{1}}$Department of Physics, University of Virginia, P. O. Box 400714,
Charlottesville, Virginia 22904-4714, USA\\
$^{{2}}$Department of Physics and Astronomy, University of Kentucky,
Lexington, Kentucky 40506-0055, USA}

\begin{abstract}
We compute the generic mode sum that quantifies the effect on  the spectrum of a harmonic field when a spherical shell is inserted into vacuum.  This encompasses a variety of problems including the Weyl spectral problem and the Casimir effect of quantum electrodynamics.  This allows us to resolve several long-standing controversies regarding the question of universality of the Casimir self-energy; the resolution comes naturally through the connection to the Weyl problem.  Specifically we demonstrate that in the case of a scalar field obeying Dirichlet or Neumann boundary conditions on the shell surface the Casimir self-energy is cutoff-dependent while in the case of the electromagnetic field perturbed by a conductive shell the Casimir self-energy is universal.  We additionally show that an analog non-relativistic Casimir effect due to zero-point magnons takes place when a non-magnetic spherical shell is inserted inside a bulk ferromagnet.
\end{abstract}


\maketitle

\section{Introduction}

Many physics problems require the evaluation of the sum
\begin{equation}
\label{sum 2 }
\mathcal{Z}(Q)=\sum_{\alpha}F(q_{\alpha};Q)\approx \int_{0}^{\infty}F(q;Q)G(q)dq
\end{equation}
where $q_{\alpha}$ is the wavevector spectrum for a harmonic field confined to a region and $F(q;Q)$ is a function that goes to zero for $q\gg Q$ monotonically and fast enough that the mode sum converges (the essence of the second step in (\ref{sum 2 }) will be explained shortly). The parameter $Q$ is the cutoff wavevector whose physical meaning depends on the system in question.  When the function $F$ is the contribution of a mode to a thermodynamic property of a harmonic field of wavevector $q$ such as the entropy, energy, or free energy, the corresponding sums (\ref{sum 2 }) describe the thermodynamics of black-body radiation in a cavity or, equivalently, the thermodynamics of a harmonic solid \cite{LL5}.  Here the Planck distribution enters the function $F(q;Q)$ guaranteeing the convergence of the sum (\ref{sum 2 }) and the temperature $T$ serves as the wavevector cutoff $Q$. For a macroscopic region of typical size $a$ the condition $T\gg 1/a$ is satisfied and there are many terms in the $1/a < q_{\alpha} < T$ range contributing to the mode sum.   

The function $F$ can also represent the zero-point energy of a harmonic field with linear dispersion law; however it is useful to amend the dependence $F=q/2$ at large wavevectors so that $F$ goes to zero at $q$ large.  This assures that the sum (\ref{sum 2 }) converges, and helps classify and resolve the divergences at high wavevectors that might otherwise occur; physically this means that a material boundary surrounding the region in question becomes invisible to short wavelength radiation \cite{Casimir,Power} and thus does not perturb the spectrum at highest wavevectors.  In this example the cutoff wavevector $Q$ is a property of the material of which the boundary is constructed.  Mode sums like (\ref{sum 2 }) are also encountered in Fermi systems \cite{LL5}.  

In a macroscopic system the discrete spectrum can be approximated by a continuous one and characterized by a density of states (DOS) $G(q)$ so that the mode sum can be represented as an integral \cite{LL5} as shown in the second step in (\ref{sum 2 }). 
The function $G(q)$, to be referred to as the Weyl DOS, represents the continuum approximation to the exact DOS $\mathcal{G}(q)=\sum_{q}\delta(q-q_{\alpha})$;  it plays a central role in the physics of finite-sized systems. The significant feature of this replacement is that $G$ is independent of $F$: it encodes the physics of the field and the boundary condition, while $F$ is the specific aspect of the system that is being studied.  

The large-$q$ behavior of the Weyl DOS $G(q)$ can be inferred from the behavior of $\mathcal{Z}(Q)$ for large cutoff by inverting the integral transform (\ref{sum 2 }).  These leading terms, known as the Weyl series \cite{Weyl}, have geometric interpretation. The Weyl series is the origin of the divergences that occur in attempts to calculate the Casimir effect \cite{Casimir,CasReviews} without use of a cutoff;  such divergences can represent real effects of the presence of the physical cutoff $Q$ \cite{BB,BD,DK,D,Schaden,Fulling10,KSLZ}.  Even though the connection between the Weyl problem and the Casimir problem has been noted previously, numerous investigations calculating the Casimir energy \cite{Bender,Lese,NP,NP2,NLS,Cog,Milton_book} seem to treat the cutoff-dependent terms as mathematical artifacts.  The latter, in a multitude of cases, can be hidden by a regularization procedure.  However there is a class of geometries (spherical shells in even number of dimensions, separate contributions of interior and exterior modes) when regularization schemes fail, and the result is presented in a formally divergent form obscuring its physical meaning.  As can be seen from our discussion, all cutoff-dependent  terms have real physical meaning explainable in terms of the Weyl problem.  

We must mention at this point that an objection against the connection between the Weyl and Casimir problems was put forward by Candelas \cite{C};  below by analysis of the general spectral problem (\ref{sum 2 }) that includes both the Weyl and Casimir problems as special cases, Candelas's objection is refuted.

Since the Weyl DOS $G(q)$ is independent of the cutoff procedure, it can be extracted from the mode sum calculated using any convenient function $F(q)$. Kac \cite{Kac} and Stewartson and Waechter \cite{SW} demonstrated the utility of the Gaussian function $F=\exp(-q^{2}/Q^{2})$ and applied it to the case of a two-dimensional region.  However the exponential $F=\exp(-q/Q)$ and power-law $F=q^{-s}$ choices are just as good.  The latter is employed in the zeta function regularization method \cite{zeta}; there is no cutoff scale  $Q$, and one studies instead the role played by the parameter $s$. 

The goal of this paper is to demonstrate how to compute the mode sum for generic $F(q)$ and thus to infer the \textit{change} in the DOS due to the insertion of  a three-dimensional spherical shell into vacuum.  This will lead to some general observations, which we believe are relevant to other geometries than the spherical shell: (1) the geometric terms in the Weyl DOS are the origin of the cutoff-dependent parts of the Casimir energy, and thus are real physical properties of the problem that cannot be regulated away; (2) cutoff-dependent contributions to the Casimir energy are not present for the case of the electromagnetic field (but generally present for problems involving a scalar field); (3) the change in the DOS for the electromagnetic case vanishes faster than any power of $1/q$.  Some of these claims have been made previously, but have not been fully accepted. 

\section{One-dimensional geometry}

To illustrate the concepts we begin with an example of a one-dimensional interval of length $a$ and a scalar harmonic field satisfying the Dirichlet boundary conditions at the interval ends. The spectrum is given by $q_{n}=\pi n/a, n=1,2,...$.  Then employing the Euler-Maclaurin summation formula \cite{EM1}
the mode sum (\ref{sum 2 }) can be transformed into
\begin{eqnarray}
\label{ Sum 3}
\mathcal{Z}(Q,a)=\sum_{n=1}^{\infty}F(\frac{\pi n}{a})&=&\frac{a}{\pi}\int_{0}^{\infty}F(q)dq-\frac{F(0)}{2}\nonumber\\
&-&\frac{\pi F'(0)}{12a}+\frac{\pi^{3}F^{(3)}(0)}{720a^{3}}-...
\end{eqnarray} 
assuming that $F(y)$ is not singular at the origin and that it and all of its derivatives vanish at infinity.  In view of Eq.(\ref{sum 2 }) this is consistent with the generalized Weyl expansion of the form
\begin{equation}
\label{ 1d_DOS}
G(q)=\frac{a}{\pi}-\frac{\delta(q)}{2}+\frac{\pi \delta'(q)}{12a} -\frac{\pi^{3}\delta^{(3)}(q)}{720a^{3}}+...
\end{equation}
with the understanding that the delta-function and its derivatives are only used as a shorthand to indicate that the mode sum is only sensitive to aspects of the long-wavelength part of $F(q)$.  

In this example the first two terms of the Weyl expansion (\ref{ 1d_DOS}) have geometrical relevance:  the leading term is proportional to the one-dimensional "volume" $a$;  the corresponding term of the mode sum (\ref{ Sum 3}) is the only one that requires a cutoff.  The next order delta-function term represents the effect of the edges of the interval.   Our expression (\ref{ 1d_DOS}) contains more terms than usually kept \cite{Weyl};  typically one keeps only the geometric terms and deals with the rest separately.  Our attitude here is that the DOS can have as many terms as needed as long as for any physical quantity the outcome can be presented in the integral form like in Eq.(\ref{sum 2 }). 

The coefficient of the second term in (\ref{ Sum 3}) (which gives rise to the delta-function term in (\ref{ 1d_DOS})) has a special place within the theory because it does not depend on the wavevector cutoff $Q$ or on the macroscopic length scale $a$;  it is just due to the presence of the boundary.  We will call this coefficient the Kac number $\mathcal{K}$; apparently this number was first computed (in any context) by Kac \cite{Kac} who found that $\mathcal{K}^{(\mathcal{D})}_{d=2}=1/6$ for a simply-connected two-dimensional region enclosed by a smooth Dirichlet curve.  For the one-dimensional Dirichlet interval Eqs.(\ref{ Sum 3}) and (\ref{ 1d_DOS}) imply that $\mathcal{K}^{(\mathcal{D})}_{d=1}=-1/2$.  The Kac number gives geometrical information about the boundary and its topology. The Kac term does not contribute into the zero-point energy, but it has other measurable consequences because $\mathcal{K}$ reflects the change in the number of states due to introduction of a boundary.  Then the classical equipartition theorem \cite{LL5} implies that the energy of a region contains a universal $\mathcal{K}T$ term \cite{BD}; in one dimension the $-T/2$ piece in the energy is the leading finite-size contribution.  

The derivative terms in (\ref{ Sum 3}) and (\ref{ 1d_DOS}) also have a special place within the theory because they only depend on the macroscopic length scale $a$ and do not depend on the cutoff.   They are responsible for the Casimir effect and its generalizations.  The generalized Casimir effect will be defined as a change of the value of the mode sum as a result of the introduction of a boundary or boundaries.  For example, the change of the mode sum as a result of inserting into vacuum of two Dirichlet points separated by a distance $a$ will be given by
\begin{equation}
\label{ sum_1d_shell}
\triangle\mathcal{Z}(Q,a)=-F(0)-\frac{\pi F'(0)}{12a}+\frac{\pi^{3}F^{(3)}(0)}{720a^{3}}-...
\end{equation}
because the mode sum for the vacuum contains only the leading term of (\ref{ Sum 3});  the first term is  due to the insertion of two Dirichlet points.  Choosing $F(q\rightarrow 0)\rightarrow q/2$ we recover the well-known $-\pi/24a$ Casimir attraction \cite {Boyer03} given by the second term of (\ref{ sum_1d_shell}).  We also note that if there were a definite function $F(q)$ that we were studying which had a nonzero third derivative at zero wavevector, Eq.(\ref{ sum_1d_shell}) would describe the consequences which would be an example of generalized Casimir effect.  Since the mode sum (\ref{ sum_1d_shell}) only contains odd derivatives of $F$, the generalized Casimir effect is absent for any function that for small wave vectors vanishes as an even power of the wavevector.  Specifically, this rules out the possibility of the Casimir effect with "non-relativistic" dispersion law $F(q\rightarrow0)\propto q^{2}$;  the same conclusion holds in the parallel plane geometry in three spatial dimensions \cite{Milton_book,Fulling03}.  However below we will demonstrate a possibility of a non-relativistic Casimir effect in spherical shell geometry.   

The mode sum (\ref{ sum_1d_shell}) is consistent with the Weyl expansion of the form
\begin{equation}
\label{ 1d_DOS_shell}
\triangle G(q)=-\delta(q)+\frac{\pi \delta'(q)}{12a}-\frac{\pi^{3}\delta^{(3)}(q)}{720a^{3}}+...
\end{equation} 
Since the original Dirichlet interval can be viewed as a one-dimensional sphere while the configuration with two Dirichlet points represents a one-dimensional spherical shell (both of radius $a/2$) it is instructive to compare the expression for the DOS (\ref{ 1d_DOS_shell}) accumulating the effect of the field modes both inside and outside of the Dirichlet interval with that given by Eq.(\ref{ 1d_DOS}) which only includes the effect of the interior modes.  We then observe that the geometric part of the DOS given by the first two terms of (\ref{ 1d_DOS}) and by the first term of (\ref{ 1d_DOS_shell}) is the sum of local effects:  the bulk $a/\pi$ term present in the original DOS (\ref{ 1d_DOS}) is cancelled between the interior and exterior modes while the effect of the edges given by the delta-function term in (\ref{ 1d_DOS}) is doubled.  This is not true of the Casimir term, which is a global property \cite{D,KSLZ}.  

\section{Three-dimensional geometry}
  
We now proceed to a calculation of the counterparts of Eqs.(\ref{ sum_1d_shell}) and (\ref{ 1d_DOS_shell}) for a spherical shell in three dimensions starting with the scalar problem which is technically more involved than its electromagnetic counterpart. 

\subsection{Scalar problem}

Special cases of the general mode sum have been considered previously.

* The sum (\ref{sum 2 }) over the modes of a scalar field inside a Dirichlet sphere was first calculated by Waechter \cite{Waechter}, using the choice $F(q)=\exp(-q^{2}t)$ ($Q\equiv 1/\sqrt{t}$). The geometric parts of the DOS for a Dirichlet shell then can be extracted by use of the additive property of the geometric parts of the DOS.  Below we will calculate the generic mode sum (\ref{sum 2 }) and the DOS directly in the shell geometry and find some difference from Waechter's results.  

* The mode sum for the case $F=q /2$ corresponds to the zero-point energy of a scalar field.  The change in the zero-point energy resulting from the imposition of a new boundary is the scalar Casimir effect.  This was first studied for the case of a spherical shell in three dimensions by Bender and Milton \cite{Bender}.  The leading terms of the Weyl DOS give rise to ultraviolet divergences which were eliminated by means of a regularization technique.  In order to study these divergences we will introduce a cutoff, so that $F = (q/2)\mathcal{C}(q/Q)$ (where $\mathcal{C}(q/Q)$ is small for large argument).  The divergences become cutoff-dependent terms; we will argue that these terms have physical meaning.

\subsubsection{Contour integral representation of the mode sum}

In spherical geometry the modes are not spaced regularly in wavevector, and their determination entails solving a transcendental equation.  We can make use of this equation to represent the mode sum as a contour integral. If $q_{n}$ and $q_{p}$ are the zeros and poles, respectively, of a function $\varphi(q)$, the sum of the values of a function $F(q)$ over these sets can be calculated by 
\begin{equation}
\label{ argument_principle}
\frac{1}{2\pi i}\oint_{C} F(q)\frac{d}{dq}\ln\varphi(q)dq=\sum_{q_{n}}F(q_{n})-\sum_{q_{p}}F(q_{p})
\end{equation}   
where the contour $C$ encloses the values being included in the sum; 
the function $F(q)$ must be analytic inside the contour \cite{contour,Barash}, and we assume that it is very small where the contour cuts the real $q$ axis.

We wish to calculate the change in the mode sum arising from the introduction into a previously empty space of a spherical shell of radius $a$ that imposes Dirichlet boundary conditions.  This will be written in the form
\begin{equation}
\label{sum_of_partial_sums}
\triangle \mathcal{Z}(Q,a)=\sum_{l=0}^{\infty}2\nu\sum_{k}(F(q_{l,k})-F(\overline{q_{l,k}}))=\sum_{l=0}^{\infty}\nu \mathcal{Z}_{\nu}(Q,a)
\end{equation}
where the factor $2\nu=2l+1$ accounts for mode degeneracy, $q_{l,k}$ and $\overline{q_{l,k}}$ are the spectra of the system with and without the shell, $l$ labels the relevant spherical harmonic and the order of the corresponding Bessel function, and $k$ labels the successive modes for given $l$.  

To use the contour representation (\ref{ argument_principle}) to calculate the partial sum $\mathcal{Z}_{\nu}(Q,a)$ we construct the function $\varphi(q)$ that vanishes when $q$ is a mode wavevector of the modified system and has poles when $q$ is a mode wavevector of the unperturbed system.
To avoid dealing immediately with the complications of a continuous spectrum, consider the modes of a scalar field interior to a large sphere of radius $b$ (with Dirichlet boundary conditions on the sphere), and the change in this spectrum caused by the introduction of a concentric Dirichlet shell of radius $a < b$.  Then
\begin{equation}
\label{argument}
\varphi_{\nu}(q) = \pi q a [J_{\nu}(qb)H^{(1)}_{\nu}(qa)-J_{\nu}(qa)H^{(1)}_{\nu}(qb)]\frac{J_{\nu}(qa)}{J_{\nu}(qb)}
\end{equation}
where $J_{\nu}(z)$ and $H^{(1)}_{\nu}(z)$ are the Bessel and Hankel functions, respectively \cite{EM1}.  The wavevectors of the modes interior to the inner sphere are the zeros of the factor $J_{\nu}(qa)$, and similarly the modes of the
unpartitioned sphere are the zeros of $J_{\nu}(qb)$ (and the poles of $\varphi$); the Bessel function combination vanishes when $q$ is the wavevector for a mode confined to the region between the two spheres. The factor $\pi qa$ has been introduced so that $\varphi$ approaches unity for large imaginary $q$.  This introduces a spurious mode at $q = 0$, which should be ignored.

When $q$ has an imaginary part and $b >> a$, $J_{\nu}(qb)$ is exponentially large and $H^{(1)}_{\nu}(qb)$ is exponentially small,  so that $\varphi$  reduces to $\pi aq J_{\nu}(qa)H^{(1)}_{\nu}(qa)$.  In this way the limit $b \rightarrow \infty$ can be taken. The net effect is the replacement 
$[J_{\nu}(qb)H^{(1)}_{\nu}(qa)-J_{\nu}(qa)H^{(1)}_{\nu}(qb)]/J_{\nu}(qb)\rightarrow H^{(1)}_{\nu}(qa)$.

Setting $q = i \nu y/a$, the partial sum $Z_{\nu}(Q,a)$ now has the form 
\begin{equation}
\label{ contour_integral_partial_sum}
\mathcal{Z}_{\nu}=\oint_{\Im y > 0} \frac{F(i \nu y/a)}{\pi i} \frac{d}{dy}\ln \left (2\nu y I_{\nu}(\nu y)K_{\nu}(\nu y) \right ) dy
\end{equation}
The original integration contour $C$ encloses the positive real $q$ axis (but not including the origin, to exclude the spurious mode) in the counterclockwise direction. We can take it to enclose the whole right half-plane $\Re q > 0$, which in the new variables is the upper half-plane $\Im y > 0$.  The part of the contour that lies at large $\Im y$ makes no contribution because the cutoff function is very small there.  This is similar to the starting point of others \cite{NP}, but we have left this in the form of a contour integral.   We observe that the function $F(q)$, which goes to zero for large real $q$,  becomes $F(i y \nu/a)$ on the imaginary axis, and is not necessarily a "cutoff" function anymore; in particular, the exponential function has constant magnitude.   For this reason it is important to keep the whole contour. 

We introduce the uniform asymptotic expansion of Debye \cite{AS} valid for $\nu \gg 1$:
\begin{eqnarray}
\label{Debye}
\ln[2\nu(1&+&y^{2})^{1/2}I_{\nu}(\nu y)K_{\nu}(\nu y)]= \frac{1}{8\nu^{2}} \Big\{\frac{1}{1+y^{2}}\nonumber\\
&-& \frac{6}{(1+y^{2})^{2}} + \frac{5}{(1+y^{2})^{3}} \Big\} + R(y,\nu)
\end{eqnarray}
The remainder function $R(y,\nu)$ is of order $\nu^{-4}$ for large $\nu$ and of order $y^{-4}$ for large $y$. It gives a finite contribution to (\ref{sum_of_partial_sums}) and thus may be disregarded while we consider the more problematic leading terms. 
With this approximation 
\begin{eqnarray}
\label{analytic_continuation1}
\mathcal{Z}_{\nu}&\approx&\oint_{\Im y>0}\frac{F(i\nu y/a)}{\pi i}\frac{d}{dy}\Big [\ln\frac{y}{\sqrt{1+y^{2}}}
+ \frac{1}{8\nu^{2}} \Big\{\frac{1}{1+y^{2}}\nonumber\\
&-&\frac{6}{(1+y^{2})^{2}} + \frac{5}{(1+y^{2})^{3}} \Big\}  \Big ]dy
\end{eqnarray}
We observe that left-hand side of Eq. (\ref{Debye}) has a branch cut along the imaginary $y$ (real $q$) axis, while the only singularity of the integrand of Eq.(\ref{analytic_continuation1})  inside the contour is at $y = i$ (that is, at $q = \nu/a$ in terms of the original variables); it is a superposition of poles of multiplicity $1$, $2$, and $3$. This gives a good approximation to the integrand for the relevant case that $y$ is real.  Some algebra leads to
\begin{eqnarray}
\label{ sum 4}
&&\triangle\mathcal{Z}(Q,a)=-\sum_{l=0}^{\infty}\nu F(\frac{\nu}{a})\nonumber\\
 &+& \frac{1}{64a} \sum_{l=0}^{\infty}[-F'(\frac{\nu}{a})
+\frac{9\nu}{a}F''(\frac{\nu}{a}) +\frac{5\nu^{2}}{a^{2}}F'''(\frac{\nu}{a})]+...\nonumber\\
\end{eqnarray}
where the primes indicate differentiation with respect to the argument of $F$.  

Similar to the one-dimensional example studied above, the mode sum (\ref{ sum 4}) can be understood with desired accuracy with the help of appropriate generalization of the Euler-Maclaurin summation formula \cite{EM1}
\begin{equation}
\label{EM2}
\sum_{l=0}^{\infty}f(l+\frac{1}{2})=\int_{0}^{\infty}f(x)dx + \frac{1}{24}f'(0)-\frac{7}{5760}f'''(0)+...
\end{equation}
with the result
\begin{equation}
\label{ sum_3d_shell}
\triangle\mathcal{Z}(Q,a)=-a^{2}\int_{0}^{\infty}F(q)qdq-\frac{F(0)}{24}+\frac{17F''(0)}{1920a^{2}}+...
\end{equation}
where we restricted ourselves to the derivative terms whose order does not exceed the second. 

\subsubsection{Analysis of the mode sum}

The result (\ref{ sum_3d_shell}) is consistent with a change in the DOS of the form
\begin{equation}
\label{ 3d_DOS_shell}
\triangle G(q)=-a^{2}q-\frac{\delta(q)}{24}+\frac{17\delta''(q)}{1920a^{2}}+...
\end{equation}
There is no bulk term proportional to $a^{3}$ in (\ref{ sum_3d_shell}) and (\ref{ 3d_DOS_shell}) because the volume of the system is not changed by the insertion of the shell.    Even though approximations were made, the first two terms of (\ref{ 3d_DOS_shell}) are exact: the second of the sums in (\ref{ sum 4}) only contributes to the last term of (\ref{ sum_3d_shell}) which also receives contribution from the first sum in (\ref{ sum 4}). Consistent with Weyl's expansion, the leading terms of (\ref{ sum_3d_shell}) have geometrical interpretations. 

The first term is proportional to the area of the shell.  Waechter \cite{Waechter}, considering only the modes inside the shell, obtained a result half as large (in our evaluation, the shell perturbs the modes on both sides; the results are in agreement).  This term has played a role in previous treatments of the Casimir energy:  when $F(q)=q/2$ for all $q$, the mode sum (\ref{ sum 4}) becomes a divergent expression.  In previous publications \cite{Bender,NP} it was evaluated (to zero!) by employing the zeta regularization technique. We will discuss this approach below.

The Kac number (the coefficient of $\delta(q)$) is $\mathcal{K}_{d=3}^{(\mathcal{D})}=-1/24$; it is negative because the modes of wavelength larger than $a$ have been suppressed by the introduction of the shell.  Modes inside and outside the shell are equally affected, so that a calculation that considers only the modes inside the sphere would give a Kac term that is half as large \cite{Kennedy}.  Waechter's calculation \cite{Waechter} overlooks the Kac number.
 
We observe that the first derivative term is absent from Eqs.(\ref{ sum_3d_shell}) and (\ref{ 3d_DOS_shell}) which is an indication that the cutoff-independent part of the Casimir effect has its origin in the remainder term $R(y,\nu)$.  Since this term  does not require a cutoff, the existing treatment \cite{Bender} is adequate and will not be repeated here.  Thus the Casimir self-energy of the Dirichlet shell is given by \cite{Fulling10}
\begin{equation}
\label{Dirichlet_Casimir_energy}
\mathcal{E}^{(\mathcal{D})}=-\frac{a^{2}}{2}\int_{0}^{\infty}q^{2} \mathcal{C}(q/Q)dq+\frac{\mathcal{BM}}{a}
\end{equation}
where we wrote $F(q)=(q/2)\mathcal{C}(q/Q)$ with $\mathcal{C}(q/Q)$ representing the physical cutoff function determined by the transmission properties of the boundary and satisfying the conditions $\mathcal{C}(0)=1$ and $\mathcal{C}(\infty)=0$.  The $1/a$ dependence of the cutoff-independent term of the self-energy (\ref{Dirichlet_Casimir_energy}) is dictated by dimensional analysis, and the numerical constant $\mathcal{BM}$ was computed by Bender and Milton \cite{Bender}. The leading cutoff-dependent term of the self-energy has to be viewed as contributing $(-1/8\pi)\int_{0}^{\infty}q^{2}\mathcal{C}(q/Q)dq \simeq -Q^{3}$ into the bare coefficient of the surface tension of the shell, considered as material membrane.  
We will see below that for the special case of electromagnetism this term does not appear.  However, for the scalar field theory with Dirichlet boundary condition this will give rise to an outward stress on the sphere surface; without a cutoff, it would be an infinite stress. This realization, originally due to Deutsch and Candelas \cite{D,C}, was recently re-expressed by Graham and co-workers \cite{Jaffe} and by Barton \cite{Barton}.  The implication is that in an experimental situation, a curved boundary might give rise to a cutoff-dependent contribution to the physically measurable Casimir stress \cite{critique}.  However, it should be noticed that a cutoff as large as $Q = 10^8$ $m^{-1}$ (equivalent to the UV for light) would only give a stress of order $0.01 N/m$, to be compared with $0.5 N/m$ for liquid mercury \cite{CRC}: the apparent divergence comes from taking the mathematics too seriously.  

Neumann boundary conditions for the scalar field can be discussed in the same way.  The boundary condition replaces $ I_{\nu} (y)$ by $\sqrt{y} d/dy (\sqrt {y} I_{\nu}(y))$ and $K_{\nu}(y)$ by $\sqrt{y} d/dy (\sqrt {y} K_{\nu}(y))$, and leads to an expression for the change in the density of states similar to Eq.(\ref{ 3d_DOS_shell})
\begin{equation}
\label{ 3d_DOS_shell_Neumann}
\triangle G(q)=a^{2}q+\frac{7\delta(q)}{24}-\frac{97\delta''(q)}{1920a^{2}}+...
\end{equation}
Since the effect of Neumann boundary conditions is to remove a constraint (continuity of the field across the shell) rather than to add one, every term has the opposite sign relative to the Dirichlet case.  Similar to the Dirichlet case the Casimir self-energy is dominated by the cutoff-dependent contribution proportional to the area of the sphere. 

\subsubsection{Non-relativistic Casimir effect}
 
Since Eqs.(\ref{ sum_3d_shell}) and (\ref{ 3d_DOS_shell}) contain second-order derivative terms, they demonstrate the possibility of a non-relativistic Casimir effect.  Since the non-relativistic Casimir effect of quantum electrodynamics is exponentially suppressed by the rest energy of elementary particles, here we discuss its condensed matter analog.  Even though the most interesting universal part of the effect, as explained below, is small compared to its cutoff-dependent piece, the calculation is of interest as a proof of principle and because the overall effect is not small.

Spin waves in ferromagnets are zero chemical potential Bose excitations having a dispersion law whose long-wavelength limit is "non-relativistic", $\omega=\gamma q^{2}$ \cite{LL9}.  Then the zero-point energy per mode is $\omega/2=\gamma q^{2}/2$.  Let us now assume that there is a non-magnetic spherical surface of radius $a$ (impenetrable to magnons) embedded inside bulk ferromagnet.  The change in the zero-point energy due to the presence of the sphere can be found by substituting  $F(q)= (\gamma q^{2}/2)\mathcal{C}(q/Q)$ into Eq.(\ref{ sum_3d_shell})
\begin{equation}
\label{nr_Casimir_energy}
\mathcal{E}^{(\mathcal{D})}_{nr}=-\frac{a^{2}\gamma}{2}\int_{0}^{\infty}q^{3} \mathcal{C}(q/Q)dq+\frac{17\gamma}{1920a^{2}}
\end{equation} 
As in previously studied cases, the effect is dominated by the leading cutoff-dependent term reducing the bare coefficient of surface tension of the surface by an amount of the order $\gamma Q^{4}$.   The sub-leading term manifests itself as a universal finite-size correction with the $\gamma/a^{2}$ dependence dictated by dimensional analysis.  The higher-order terms of the Debye expansion do not contribute anything further.  

\subsubsection{Connection to the zeta function regularization method}

The zeta function regularization method evaluates the mode sum for the choice $F(q) = q^{-s}$.  Its connection to the physical cutoff approach based results has a large literature \cite{zetacutoff};  below this relationship is discussed in the spherical shell geometry.  For sufficiently large $s$, the sums and integrals are convergent at large $q$; the result, referred to as the spectral zeta function, is analytically continued to physically relevant $s$.  The relativistic and non-relativistic Casimir energies and the Kac number are the $s = -1$, $s=-2$ and $s=0$ cases, respectively.   For example, consider the evaluation of the sum (\ref{ sum 4}). With the choice $F(q)=q^{-s}$, this becomes
\begin{eqnarray}
\label{modesum5}
\triangle\mathcal{Z}(s,a) &=& - a^{s} \zeta \left (s-1, \frac{1}{2}\right )\nonumber\\
 &-& \frac {s^2 a^{s}}{64}(5 s + 6)\zeta \left (s+1, \frac{1}{2}\right )+ ...\nonumber\\
 &=&-a^{s}(2^{s-1}-1)\zeta(s-1)\nonumber\\
 &-&\frac{s^{2}a^{s}}{64}(5s+6)(2^{s+1}-1)\zeta(s+1)+...
\end{eqnarray}
where $\zeta(x,y)$ and $\zeta(x)$ are the Hurwitz and Riemann zeta functions, respectively \cite{AS}.  Although the sum is not convergent for $s < 2$, the zeta functions are defined by analytic continuation, and for physically interesting cases the spectral zeta function (\ref{modesum5}) reproduces all previously found universal results.  Indeed, employing $\zeta(-1)=-1/12$, $\zeta(-2)=0$, and $\zeta(-3)=1/120$ \cite{AS} we find $\triangle\mathcal{Z}(-1,a)=0$ (for the ordinary relativistic Casimir energy), $\triangle\mathcal{Z}(-2,a)=17/960a^{2}$ (for the $\gamma/2$ coefficient of the non-relativistic Casimir energy in (\ref{nr_Casimir_energy})), and  $\triangle\mathcal{Z}(0,a)=-1/24$ (for the Kac number appearing in Eqs.(\ref{ sum_3d_shell}) and (\ref{ 3d_DOS_shell})).  

The cutoff-dependent terms are also represented in the spectral zeta function.  Eq. (\ref{modesum5}) has a pole of residue $-a^{2}$ at $s=2$ which corresponds to the leading cutoff-dependent term in the mode sum (\ref{ sum_3d_shell}).  Indeed, evaluating the latter with the choice $F=q^{-s}$ and lower integration limit $q=\epsilon$ (because there is a lowest wavevector mode) we find $-a^{2}\epsilon^{2-s}/(s-2)$ which is a pole of residue $-a^{2}$ at $s=2$ in agreement with the pole structure of (\ref{modesum5}).   We claim that in general the cutoff-dependent terms in a mode sum are represented by the poles of the spectral zeta function that have to be bypassed on the way from $s > 2$ (or wherever the sums and integrals actually converge) to the physical case that is being studied ($s=-1$, $s=-2$, or $s=0$ in the above examples).  Specifically, a pole of residue $A$ at $s=\sigma$ signals the presence of an $Aq^{\sigma-1}$ piece in the DOS which via (\ref{sum 2 }) implies the presence of cutoff-dependent contribution into the mode sum.   We will insist that the cutoff-dependent terms have physical meaning; they are a part of the Casimir energy (or whatever property is being studied) that is larger than the cutoff-independent contribution that is usually calculated.  

This connection explains why the zeta regularization fails for the calculation of the scalar Casimir effect due to a Dirichlet circle \cite{Bender}.  The value of the Casimir term should be given by the case $s = -1$ for the spectral zeta function $\triangle Z(s)$; however it has a pole there.  The existence of this pole is directly related to the logarithmically divergent $\triangle Z(Q)$ found by Sen \cite{Sen}. 

With this understanding, the zeta regularization is a powerful method for evaluating the mode sum.  For example, suppose that (for the case of a spherical Dirichlet shell introduced into a scalar field) we wished to evaluate the change in the mode sum for the function $F(q) = q^{1/2}$, which is the case $s=-1/2$ of the general problem $F(q) = q^{-s}$.  The singular behavior at $q = 0$ is not consistent with the assumptions of the Euler-Maclaurin expansion, so (\ref{ 3d_DOS_shell}) does not apply.  Since going from large $s$ to the case of interest, $s=-1/2$, one encounters a pole at $s=2$, there exists a corresponding cutoff dependent contribution.  As above we will have to introduce a physical cutoff function ${\mathcal C}$.  The value of (\ref{modesum5}) at $s = -\frac {1}{2}$ is the cutoff-independent part. 
Thus for this example
\begin{eqnarray}
\label{modesum6}
&&\triangle\mathcal{Z}(Q,a) = - a^{2} \int q^{3/2} {\mathcal C}(q/Q) dq
\nonumber
\\
&-& a^{-1/2} \left ((2^{-3/2}-1)\zeta(-\frac {3}{2}) + \frac {7 (\sqrt{2}-1)}{512}\zeta(\frac {1}{2})\right )
\end{eqnarray}

\subsection{Electromagnetic problem}

The most relevant mode sum problems involve finding the counterparts to Eqs.(\ref{ sum_3d_shell}) and (\ref{ 3d_DOS_shell}) for the case of electromagnetic vacuum perturbed by a conductive shell.  

\subsubsection{Background}

Boyer \cite{Boyer68} demonstrated that the Casimir energy for a spherical shell varies as $1/a$ and gives an outward stress on a sphere; his result is cutoff-independent.    Boyer's finding has been confirmed in several complementary  calculations \cite{BD,MDS,NP}. 

The Weyl and Casimir problems for the case of electromagnetic field were considered by Balian and Bloch and by Balian and Duplantier \cite{BD,BB}. Their conclusion was that the Casimir energy for an arbitrarily shaped perfectly conducting shell in the electromagnetic case is in general cutoff-independent: the differing boundary conditions for electric and magnetic fields give canceling contributions to the leading terms of the mode sum, and thus is a special property of the vector character of the electromagnetic field.  They also showed that the Kac number for this problem can be calculated as a surface integral over the local curvatures $\kappa_{1,2}$ 
\begin{equation}
\label{BD}
\mathcal{K}=\frac{1}{128\pi}\int d\sigma (3\kappa_{1}^{2}+3\kappa_{2}^{2}+2\kappa_{1}\kappa_{2})-\textsl{n}
\end{equation}
where $\textsl{n}$ is the genus of the surface, which can also be written as an integral of the surface curvature \cite{GB}:
\begin{equation}
\label{GB}
1- n = \frac{1}{4\pi} \int d\sigma \kappa_{1}\kappa_{2} .
\end{equation}
For the sphere this gives $\mathcal{K}=1/4$. In the high-temperature limit the energy of a conductive shell is given by $\mathcal{K}T$ as a consequence of the equipartition theorem. 

However, Candelas \cite{C} reanalyzed the problem of the conducting shell and argued that there is a cutoff-dependent contribution to the Casimir energy of a conductive shell which cannot be explained in terms of the Weyl problem; he finds that at zero temperature there is a contribution involving the surface integral that is quadratic in the curvatures (and thus can be written in terms of $\mathcal{K}$ and $\textsl{n}$). Below,  by computing the generic mode sum which encompasses both the Weyl and Casimir problems, we settle the controversy by refuting the statement of Candelas \cite{C}.    

\subsubsection{Analysis of the mode sum}

A derivation of an expression for the Casimir energy of the spherical shell beginning from the contour integral representation (\ref{ argument_principle}) was given by Nesterenko and Pirozhenko \cite{NP};   its generalization to the case of the generic mode sum requires only a few changes.  Therefore we only quote counterparts of Eqs.(\ref{sum_of_partial_sums}) and (\ref{ contour_integral_partial_sum})
\begin{equation}
\label{em_sum_of_partial_sums }
\triangle\mathcal{Z}^{(\mathcal{EM})}(Q,a)=\sum_{l=1}^{\infty}\nu \mathcal{Z}_{\nu}^{(\mathcal{EM})}(Q,a)
\end{equation}
\begin{equation}
\label{em_im_axis integral}
 \mathcal{Z}_{\nu}^{(\mathcal{EM})}=\oint_{\Im y>0} \frac{F(i\nu y/a)}{\pi i} \frac{d}{dy}\left (\ln \{1-[\sigma'_{\nu}(\nu y)]^{2}\}\right )dy
\end{equation}
where 
\begin{equation}
\label{sigma_exact}
\sigma_{\nu} (y)=yI_{\nu}(y)K_{\nu}(y)
\end{equation}
and the prime in (\ref{em_im_axis integral}) indicates differentiation with respect to the argument of $\sigma_{\nu}(y)$.  If we choose $F(q)=(q/2)\exp(-q/Q)$ then Eqs.(\ref{em_sum_of_partial_sums })-(\ref{sigma_exact}) reduce to an expression for the energy analyzed by Milton, DeRaad and Schwinger \cite{MDS}.  

The subsequent analysis mirrors the steps undertaken in treating the scalar version of the problem.  In the present case the Debye expansion (\ref{Debye}) amounts to the approximation \cite{MDS}
\begin{equation}
\label{sigma_asymptotic}
[\sigma_{\nu}'(\nu y)]^{2}\approx \frac{1}{4\nu^{2}(1+y^{2})^{3}}
\end{equation}  
for $\nu$ large.   Evaluating the resulting contour integral 
\begin{equation}
\label{em_analytic_continuation2}
\mathcal{Z}_{\nu}^{(\mathcal{EM})} \approx -\oint_{\Im y>0}\frac{F(i\nu y/a)}{4 \pi i \nu^{2}}\frac{d}{dy}\left (\frac{1}{(1+y^{2})^{3}}\right )dy
\end{equation}
and substituting the outcome into (\ref{em_sum_of_partial_sums }) we find the electromagnetic counterpart of Eq.(\ref{ sum 4}) 
\begin{eqnarray}
\label{em_sum 4}
\triangle \mathcal{Z}^{(\mathcal{EM})}(Q,a)&=&-\frac{1}{32a}\sum_{l=1}^{\infty}[3F'(\frac{\nu}{a})\nonumber\\
&-&3\frac{\nu}{a}F''(\frac{\nu}{a})+\frac{\nu^{2}}{a^{2}}F'''(\frac{\nu}{a})]+...
\end{eqnarray}
We see that Eq.(\ref{em_sum 4}) does not have an analog of the first sum in (\ref{ sum 4}), indicating that  a formal ($Q =\infty$) treatment would face weaker divergences than in the scalar case.  If we choose $F(q)=q/2$ for all $q$, then Eq.(\ref{em_sum 4}) reduces to the divergent expression for the Casimir energy found by Nesterenko and Pirozhenko \cite{NP}.  They eliminated the divergence through use of the zeta function regularization method, leading to a universal $1/a$ result.  However, the evaluation is even simpler if $F(q)$ contains an ultraviolet cutoff, because then the sum is always convergent.   

To apply the Euler-Maclaurin summation formula (\ref{EM2}) to Eq.(\ref{em_sum 4}), write 
\begin{equation}
\label{manipulation}
\sum_{l=1}^{\infty}[...]\approx\sum_{l=0}^{\infty}[...]-3F'(0)
\end{equation}
Then the mode sum (\ref{em_sum 4}) becomes
\begin{equation}
\label{sum_em_shell}
\triangle \mathcal{Z}^{(\mathcal{EM})}(Q,a)=\frac{F(0)}{4} + \frac{3F'(0)}{32a} +... ,
\end{equation}
The noteworthy features of this expression, compared to its scalar counterpart (\ref{ sum_3d_shell}), are lack of the $F''(0)$ term; the possibility of taking the limit of infinite cutoff scale; and the lack of dependence on the form of the cutoff function itself.   The cutoff function did play a role, however: from the large $\nu$ and $y$ dependences of (\ref{sigma_asymptotic}) we can see that in the absence of the cutoff function, the integral (\ref{em_im_axis integral}) converges and the sum (\ref{em_sum_of_partial_sums }) diverges.  The effect of the cutoff function is to prevent the change in variables that would allow doing these calculations sequentially.

Inclusion of higher order contributions from the Debye approximation in Eq.(\ref{sigma_asymptotic}) will allow evaluation of further terms in (\ref{sum_em_shell}).  These   also allow the limit of infinite cutoff to be taken.  The implication is that for the electromagnetic problem the change in mode sum of $F(q)=q^{n} \exp(-q/Q)$ is cutoff-independent for all $n$.  This in turn means that
\begin{equation}
\label{ordern}
{\Delta\cal Z^{(\mathcal{EM})}(\infty)} =
\int_{0}^{\infty} q^{n} G(q) dq
\end{equation}
is finite for all $n$, so that the DOS must fall off faster than any power of $q$.

Candelas's treatment of this problem differs in that he treats specifically the case $F(q) = (q/2) \exp(-q/Q)$, and then  converts the contour integral into an integral over the imaginary axis.  This is valid if $Q$ is finite, but he then tries to discuss the $Q = \infty$ limit before doing the sum.   As can be seen from Eq. (\ref{em_sum 4}), this gives a divergent expression.  The origin of this failure is that the approximation Eq. (\ref{sigma_asymptotic}), though sufficiently accurate for real $y$, fails where the contour crosses the imaginary $y$ (real $q$) axis.  The original expression (in terms of Bessel functions) represents a infinite set of modes; it has a branch cut along the real axis which represents the change in the number of states (the Kac number).  So long as there is a cutoff at large wavevector, this plays no role; however, the contour integral must be performed before the limit $Q = \infty$ is taken.   

Having dismissed the cutoff on wavevector, Candelas introduces an additional cutoff on the order of the Bessel function $\nu$.  This introduces an extra $\nu$ dependence in the terms of the sum Eq. (\ref{em_sum 4}), which spoils the feature that it is a sum of exact differentials with respect to $\nu$ (regarded as a 
real variable), so that the integral term in the Euler-Maclaurin rule can no longer be evaluated.  This may be a relevant observation, because it challenges our assumption in Eq. (\ref{sum 2 }) that the cutoff depends only on wavevector.  If the cutoff function is taken to represent the transition from decoupling of interior and exterior at long wavelengths ($F \rightarrow q/2$ at small frequency) to transparency at short wavelengths ($F \rightarrow 0$ at large frequency), there should indeed be a separate dependence on $\nu$, since this represents the angle of incidence of the wave.  However, transmission through the boundary should happen more readily at normal incidence (small $\nu$), whereas the cutoff Candelas imposes has the opposite effect.  Our conclusion is that there is no cutoff-dependent term for the problem that we consider (Eq. (\ref{sum 2 })), but leave slightly open the possibility that there could be one for a real metal.  Again assuming $Q =10^{8} m^{-1}$, the energy that Candelas proposes is of order ${\mathcal K} \hbar c Q \simeq10^{-18} J$, which will be comparable to the surface energies of condensed matter origin only for objects smaller than a nanometer.  
 
The DOS is given by
\begin{equation}
\label{em_shell_DOS}
\triangle G^{(\mathcal{EM})}(q)=\frac{\delta(q)}{4}-\frac{3\delta'(q)}{32a}+...
\end{equation}   
The expressions (\ref{sum_em_shell}) and (\ref{em_shell_DOS}) do not contain a term proportional to the area of the shell,  which is a sign of zero coefficient of surface tension in the Casimir problem.  The only geometric contribution present is the Kac term with $\mathcal{K}^{(\mathcal{EM})}=1/4$.  In contrast to its scalar counterpart $\mathcal{K}_{d=3}^{(\mathcal{D})}=-1/24$ (see Eqs.(\ref{ sum_3d_shell}) and (\ref{ 3d_DOS_shell})) the electromagnetic Kac number is positive which means that a conductive  boundary \textit{increases} the number of states.  Our result for the Kac number agrees with the Balian-Duplantier prediction (\ref{BD}).

The Casimir energy is the mode sum for the case $F(q) = q/2$.  Its value $3/64a$ is determined by the derivative term of (\ref{sum_em_shell}).  The $3/64a$ universal answer agrees with the evaluation given by Nesterenko and Pirozhenko \cite{NP}, who used the zeta function regularization procedure.  Even though this represents the bulk of the Casimir energy for a conductive spherical shell \cite{MDS}, there are $1/a$ corrections to this result due to higher order terms in the asymptotic expression (\ref{sigma_asymptotic}).  These corrections do not require a cutoff and have been calculated \cite{MDS}.     

\subsubsection{Electromagnetic spectral problem in the zeta function regularization method}

By choosing $F=q^{-s}$ the electromagnetic mode sum (\ref{em_sum 4}) becomes the electromagnetic spectral zeta function
\begin{eqnarray}
\label{ EM_spectral_zeta_function}
\Delta \mathcal{Z}^{(\mathcal{EM})}(s,a)&=& \frac{sa^{s}}{32}(s+2)(s+4)\nonumber\\
 &\times& [(2^{s+1}-1) \zeta(s+1)-2^{s+1}]+...
\end{eqnarray}
which accumulates all previously found results.  First, by comparing Eq.(\ref{ EM_spectral_zeta_function}) with its scalar counterpart (\ref{modesum5}) we observe that while the latter has a pole at $s=2$ warning us of the possibility of the presence of cutoff-dependent contribution into the mode sum proportional to the sphere area,  the electromagnetic spectral function (\ref{ EM_spectral_zeta_function}) is everywhere analytic.  The consequence is that Eq.(\ref{ EM_spectral_zeta_function}) naively evaluated at physically interesting $s$ does not overlook cutoff-dependent contributions.  Indeed, all previously found results are reproduced:

(i)  for $s=-1$ (ordinary Casimir effect) we find $\Delta \mathcal{Z}^{\mathcal{(EM)}}(-1,a) = 3/32a$,  the amplitude of the $F'(0)$ term in Eq.(\ref{sum_em_shell}) as expected;

(ii)  for $s=-2$ (non-relativistic Casimir effect) we obtain  $\Delta \mathcal{Z}^{\mathcal{(EM)}}(-2,a) = 0$ consistent with the absence of the second derivative $F''(0)$ term in Eq.(\ref{sum_em_shell});

(iii)  for $s=0$ (the Weyl problem) we arrive at  $\Delta \mathcal{Z}^{(\mathcal{EM})}(0,a) = 1/4$ which is again the right answer for the Kac number, the amplitude of the $F(0)$ term in Eq.(\ref{sum_em_shell}).

Like its scalar counterpart (\ref{modesum5}), the electromagnetic spectral function (\ref{ EM_spectral_zeta_function}) is more informative than the Euler-Maclaurin based mode sum (\ref{sum_em_shell}):  if the function $F(q)$ is singular at $q=0$, the assumptions of the Euler-Maclaurin expansion are not satisfied, and thus (\ref{sum_em_shell}) is invalid.  In such cases Eq.(\ref{ EM_spectral_zeta_function}) continues to be applicable.  

\section{Summary}

We have shown how to calculate the generic mode sum for three cases: a scalar field in one dimension perturbed by a Dirichlet boundary; a scalar field in three dimensions, perturbed by the introduction of a spherical shell; and the electromagnetic field in three dimensions, perturbed by the introduction of a conducting shell.  We have shown that these will in general contain contributions of geometric origin (the Weyl terms) which require a cutoff and have physical meaning.  We have shown how to extract the Kac number and the Casimir term when it exists.   

\section{acknowledgments}

This work was supported by US AFOSR Grant No. FA9550-11-1-0297.

\end{document}